\documentclass[twocolumn]{aastex61}

\accepted{November 29, 2017}
\submitjournal{ApJ}

\shorttitle{Ellipsoidal Variation Model Testing}
\shortauthors{Gai et al.}

\usepackage{amssymb}
\usepackage{graphicx}
\usepackage{amsmath}
\usepackage{geometry}
\usepackage{mathrsfs}
\usepackage{wasysym}
\usepackage{dcolumn}   
\usepackage{bm} 
\usepackage{tabularx}
\usepackage{slashbox}
\usepackage[english]{babel}

\DeclareGraphicsExtensions{.pdf,}

\begin{document}

\title{Bayesian Model Testing of Ellipsoidal Variations on Stars due to Hot Jupiters}

\author{Anthony D. Gai}
\email{agai@albany.edu}
\affil{Department of Physics, University at Albany, 1400 Washington Avenue, Albany, NY 12222}

\author{Kevin H. Knuth}
\affil{Department of Physics, University at Albany, 1400 Washington Avenue, Albany, NY 12222}
\email{kknuth@albany.edu}






\begin{abstract}
\addcontentsline{toc}{chapter}{\numberline{}Abstract}%
A massive planet closely orbiting its host star creates tidal forces that distort the typically spherical stellar surface. These distortions, known as ellipsoidal variations, result in changes in the photometric flux emitted by the star, which can be detected within the data from the Kepler Space Telescope. Currently, there exist several models describing such variations and their effect on the photometric flux. By using Bayesian model testing in conjunction with the Bayesian-based exoplanet characterization software package EXONEST, the most probable representation for ellipsoidal variations was determined for synthetic data and the confirmed hot Jupiter exoplanet Kepler-13A~b. The most preferred model for ellipsoidal variations observed in the Kepler-13 light curve was determined to be EVIL-MC. Among the trigonometric models, the Modified Kane \& Gelino model provided the best representation of ellipsoidal variations for the Kepler-13 system and may serve as a fast alternative to the more computationally intensive EVIL-MC. The computational feasibility of directly modeling the ellipsoidal variations of a star are examined and future work is outlined. Providing a more accurate model of ellipsoidal variations is expected to result in better planetary mass estimations.
\end{abstract}

\section{Introduction}\label{Intro}

Hot Jupiters are gas giant planets orbiting closely to their host star, many having orbital periods less than a couple of days. They represent an unusual class of exoplanet because there is no analogue within our Solar System. The transit method of exoplanet detection is well suited to both detecting and characterizing hot Jupiter exoplanets. In addition to transit effects, large planets with short orbital periods may introduce detectable photometric variations to the observed light curve.

The Bayesian-based exoplanet characterizing software package EXONEST, developed by \cite{Placek2013, Placek2014, PlacekThesis}, uses nested sampling methods to perform Bayesian model testing and focuses on modeling transits and the four most prominent photometric effects found in Kepler light curves: reflected light, thermal emission, doppler boosting, and ellipsoidal variation.

The effect due to reflected light assumes the star radiates isotropically and the planet reflects light isotropically. The total reflected light flux is computed by integrating over the illuminated portions of the planet observable from the line-of-sight. This effect has a period equal to the orbital period of the planet.

Thermal emission, which also has a period equal to the orbital period, represents the contribution of the blackbody radiation from the planet. In circular orbits, reflected light and thermal emission vary identically and cannot be disentangled. By combining observations from multiple bandpasses (e.g. Kepler and TESS), reflected light and thermal emission can be distinguished and provide improved constraints on geometric albedo, day-side temperatures, and night-side temperatures (\cite{Placek2015}). A sufficient eccentricity ($e  \gtrsim 0.3$) in the orbit breaks the degeneracy and allows one to independently identify reflected light and thermal emissions using only the Kepler light curve (\cite{Placek2013, PlacekThesis, Knuth2017}).

As the host star orbits the barycenter of the system, the star will move toward and away from the observer. Doppler boosting is a relativistic effect that simplifies roughly to an increase in observed stellar flux as the star moves toward the observer as the star orbits the system center of mass and a decrease in observed flux as the star recedes. This approximation is the combination of many smaller effects, most notably the transformation of the energy-momentum four-vector between the frame of the star and the observer (\cite{Rybicki1974}), and the shift of the stellar emission flux due to the Doppler effect within the Kepler bandpass. Generally, the transformation of the energy-momentum four-vector will create a decrease in observed flux when the star is receding from the observer and an increase when the star approaches the observer. The Doppler effect is more complicated and depends on the location of peak emission. \cite{Jackson2012} report that HAT-P-7, a bluer star, will exhibit an increase in stellar flux as the star approaches and a decrease in stellar flux when receding. Doppler boosting has a period equal to the orbital period but has a 90$^o$ phase shift compared to reflected light or thermal emission.


Ellipsoidal variation is caused by distortion of the star due to gravitational effects of a companion exoplanet. This distortion results in photometric variations consisting of two maxima per orbit, each occuring when the greatest stellar surface area is being observed. This effect may also include gravity-darkening and differences in limb-darkening due to tidal interactions, which create geometric distortions on the stellar surface. Future models may also account for geometrical deviations due to stellar rotation. 


Future missions will improve the photometric variation precision into the low ppm range. The project goal for the \textit{CHaracterising ExOPlanet Satellite (CHEOPS)} is to reach $<$10~ppm precision in flux detection with an integration time of 6 hr. The long integration time is not optimal for observing phase curve effects on hot Jupiter exoplanets. \textit{CHEOPS} is scheduled for launch in 2018 and will target systems with previously confirmed exoplanets. This mission moves beyond the goal of detection into more detailed characterization and identification. The bandpass will span between 0.4 and 1.1~microns (\cite{Broeg2013}).

The \textit{PLAnetary Transits and Oscillations (PLATO)} mission is a European Space Agency run mission which will provide long-term, short cadence $(<1~minutes)$ data at high precision. \textit{PLATO} will improve on the data collected in the Kepler mission and extend the search for habitable planets around sun-like stars (\cite{PlatoSciReq}). In addition, the project will study the effects of stellar oscillations and will provide more information about star mass, size, and age, which are all vital priors to exoplanet characterization. The target launch date is 2025 (\cite{PLATORedbook}).

In Section \ref{ModelsIntro} we introduce the time evolution description of the system and each of the models for ellipsoidal variations examined in this work. In Section \ref{Application} we examine the residual differences between the models, perform model testing on synthetic data, and apply the model testing framework to a confirmed hot Jupiter exoplanet, Kepler-13A~b. Lastly, we discuss future work in Section \ref{FutureWork} and conclusions in Section \ref{Conclusions}.

\section{Ellipsoidal Variation Models}\label{ModelsIntro}

\subsection{System Description}

The motion of a single planet orbiting a single central star is described by the solution to the Kepler problem. The resulting equation of motion describes the planet moving in an elliptical orbit,

\begin{equation}
r(t) =  \frac{a(1-e^2)}{1+e \cos \nu(t)},
\end{equation}
where $a$ is the orbital semi-major axis, $e$ is the orbital eccentricity, and the angle $\nu$ is the true anomaly.

The position of the planet as a function of time may be written in Cartesian coordinates as

\begin{equation}
\begin{pmatrix} x \\ y \\ z  \end{pmatrix} = r(t) \begin{pmatrix} \cos \big(\omega + \nu(t)\big) \\  \sin \big(\omega + \nu(t)\big )\cos i \\ \sin \big(\omega + \nu(t) \big)\sin i  \end{pmatrix}.
\end{equation}
where $\omega$ is the argument of periastron and $i$ is the orbital inclination. The line-of-sight is taken to be along the $z$-axis,  $\hat{r'}=\hat{z}$. The phase angle, which is useful when describing motion around the orbit, is defined as


\begin{equation}
\begin{split}
\theta(t) &= \arccos(\hat{r}(t)\cdot \hat{r'}) \\
            &= \arccos \Big( \frac{z(t)}{r(t)} \Big) \\
            &= \arccos \Big( \sin \big(\omega + \nu(t)\big) \sin(i)\Big) .
\end{split}
\end{equation} 

This phase angle is used when describing the photometric effects due to a planet.

\subsection{Ellipsoidal Variations}

Ellipsoidal variations were first examined in non-eclipsing binary stars. The model was later extended to describe the interaction between a star and companion exoplanet. The amplitude of the effect was described by \cite{Loeb2003} as

\begin{equation}\label{EXONEST_Ellip}
Amplitude = \beta \frac{M_P}{M_{S}} \Bigg( \frac{R_{S}}{r(t)}\Bigg) ^3,
\end{equation}
where $M_P$ is the planet mass, $M_{S}$ is the stellar mass, $R_{S}$ is the stellar radius, and $r(t)$ is the star-to-planet distance. \cite{Morris1985} derived $\beta$, the effect of linear gravity-darkening and limb-darkening, which is represented by
\begin{equation}
\beta = 0.15 ~\frac{(15+u)(1+g)}{(3-u)},
\end{equation}
where $u$ and $g$ are the limb-darkening and gravity-darkening coefficients, respectively. Limb-darkening and gravity-darkening coefficients can be estimated by modeling metallicity and effective temperature (\cite{Heyrovsky2007,Sing2010}).

\subsection{Beaming, Ellipsoid, and Reflection (BEER)}
The BEER model was developed by  \cite{Faigler2007} to describe the periodic deviations in the observed flux of a star due to a short-period planet. The model describes the effects introduced in Section \ref{Intro} as trigonometric functions. Equation \eqref{BEER_eq} is the BEER model for the flux observed from ellipsoidal variations. Since BEER approximates the effect as trigonometric functions, the algorithm runs quickly. However, the actual effect is more complicated than a simple sinusoid. BEER models the orbital motion with a time step as a fraction of the orbital period


\begin{equation}\label{BEER_eq}
\frac{F_{ellip}(t)}{F_S} \approx -\beta \frac{M_P}{M_S} \Bigg( \frac{R_S}{a} \Bigg)^3 \sin^2(i)\cos\Bigg(\frac{2\pi}{P/2 }t\Bigg),
\end{equation}
where the flux due to the ellipsoidal variations $F_{ellip}$ is normalized by the stellar flux $F_S$, $M_P$ is the planet's mass, $M_S$ is the stellar mass, $R_S$ is the stellar radius, $a$ is the star-planet distance, $i$ is the inclination, and $P$ is the orbital period of the planet. For circular orbits, the phase angle $ \theta = \frac{2 \pi}{P} t$. The relation does not hold in general for eccentric orbits. By replacing the factor $\frac{2\pi}{P}t$ in BEER with orbital phase, the effect becomes
\begin{equation}
\frac{F_{ellip}(t)}{F_S} \approx -\beta \frac{M_P}{M_S} \Bigg( \frac{R_S}{a} \Bigg)^3 \sin^2(i)\cos\Big(2\theta(t)\Big).
\end{equation}

\subsection{Kane and Gelino (2012)}
\cite{KaneGelino2012} proposed a different model also based on the amplitude of the effect described by \cite{Loeb2003}. The effect is hypothesized to be proportional to the projected separation distance, $\zeta$, between the star and planet

\begin{equation}
\begin{split}
\zeta &= \frac{\sqrt{x^2 + y^2}}{r} \\
&= \frac{\sqrt{\Big( r \cos\big(\omega+\nu(t)\big) \Big)^2 + \Big( r \sin \big(\omega + \nu(t) \big) \cos i \Big)^2}}{r}
\end{split}
\end{equation}



The estimated ellipsoidal variation from the projected separation distance simplified written alongside the amplitude from \cite{Loeb2003} is given by

\begin{equation}\label{KG2012Eq}
\begin{split}
\frac{F_{Ellip}(t)}{F_{S}} \approx  \beta \frac{M_P}{M_S} & \Big(\frac{R_S}{a}\Big)^3 \Big(\cos^2\big(\omega + \nu(t)\big) + \\
 &\sin^2\big(\omega + \nu(t)\big)\cos^2 i \Big) ^{\onehalf}.
\end{split}
\end{equation}
Substitution using the phase angle results in

\begin{equation}
\frac{F_{Ellip}(t)}{F_{S}} \approx \beta \frac{M_P}{M_S} \Big( \frac{R_S}{a} \Big)^3 \sin\big(\theta(t)\big).
\end{equation}

Unlike the BEER representation for ellipsoidal variations, which allows for both positive and negative changes in flux, the Kane \& Gelino model is strictly positive with a minimum value of zero. This model also consists of a trigonometric function and therefore will run relatively quickly. It should be noted that the model contains a discontinuity in the first derivative at the location of minimal ellipsoidal variation (see Figure \ref{fig:All_Models_Plot}). The discontinuity is not seen in simulations of ellipsoidal variations using gravitational isopotentials.

\subsection{Kane and Gelino (Modified)}

A modified version of the Kane and Gelino (2012) model, proposed by \cite{Placek2013}, introduces a square above the  trigonometric quantities in Equation \eqref{KG2012Eq}. Squaring this term is analogous to assuming that the effect is proportional to the square of the projected star-planet distance. This removes the discontinuity in the first derivative of the original model (see Figure \ref{fig:All_Models_Plot}). The equation is given by

\begin{equation}
\begin{split}
\frac{F_{Ellip}(t)}{F_{S}} \approx  \beta \frac{M_P}{M_S} &\Big(\frac{R_S}{a}\Big)^3 \Big(\cos^2\big(\omega + \nu(t)\big) + \\
&\sin^2\big(\omega + \nu(t)\big)\cos^2 i \Big),
\end{split}
\end{equation}
which simplifies using the phase angle to
\begin{equation}
\frac{F_{ellip}(t)}{F_S} \approx \beta \frac{M_P}{M_S} \Big( \frac{R_S}{a} \Big)^3 \sin^2\big(\theta(t)\big).
\end{equation}

\subsection{Ellipsoidal Variations Induced by a Low-mass Companion (EVIL-MC)}
EVIL-MC was developed by \cite{Jackson2012} to more accurately model the ellipsoidal variation. EVIL-MC was originally written in IDL and has been translated to MATLAB for compatibility with EXONEST by \cite{GaiThesis}. This model determines the shape of the star by projecting a grid onto a sphere and determining the deviation from sphericity caused by the planet for each projected stellar grid point. \cite{Jackson2012} applied this model to planets in circular orbits.

The deviation from sphericity at a point on the stellar surface due to a companion exoplanet was derived from the gravitational equipotential of the star-planet system. The deviation from sphericity on the stellar surface at a point along $\hat{R}_S$ due to a companion planet at position $\vec{A}$ relative to the center of the star is described by
\begin{equation}\label{EVILMC_dr}
\begin{split}
\delta R = q \Big([a^2 - 2a\cos\psi& +1]^{-1/2} - [a^2+1]^{-1/2}- \\
&\frac{\cos\psi}{a^2}\Big) - \frac{\omega^2}{2a^3}\cos^2\lambda,
\end{split}
\end{equation}
where
\begin{equation}
\begin{split}
q &= \frac{M_P}{M_S}\\
a &= \frac{A}{R_S},
\end{split}
\end{equation}
where $q$ is the ratio of the planet mass to the stellar mass, $a$ is the distance between the star and planet normalized by the unperturbed stellar radius,  $\cos\psi = \hat{R}_S\cdot\hat{A}$, $\omega$ is the stellar rotation rate, and $\cos\lambda = \hat{R}_S\cdot\hat{\omega}$. For this paper, the stellar rotation was set to zero, $\omega = 0$.

\begin{figure*}[htb!]
\centering
    \includegraphics[width=0.95\textwidth]{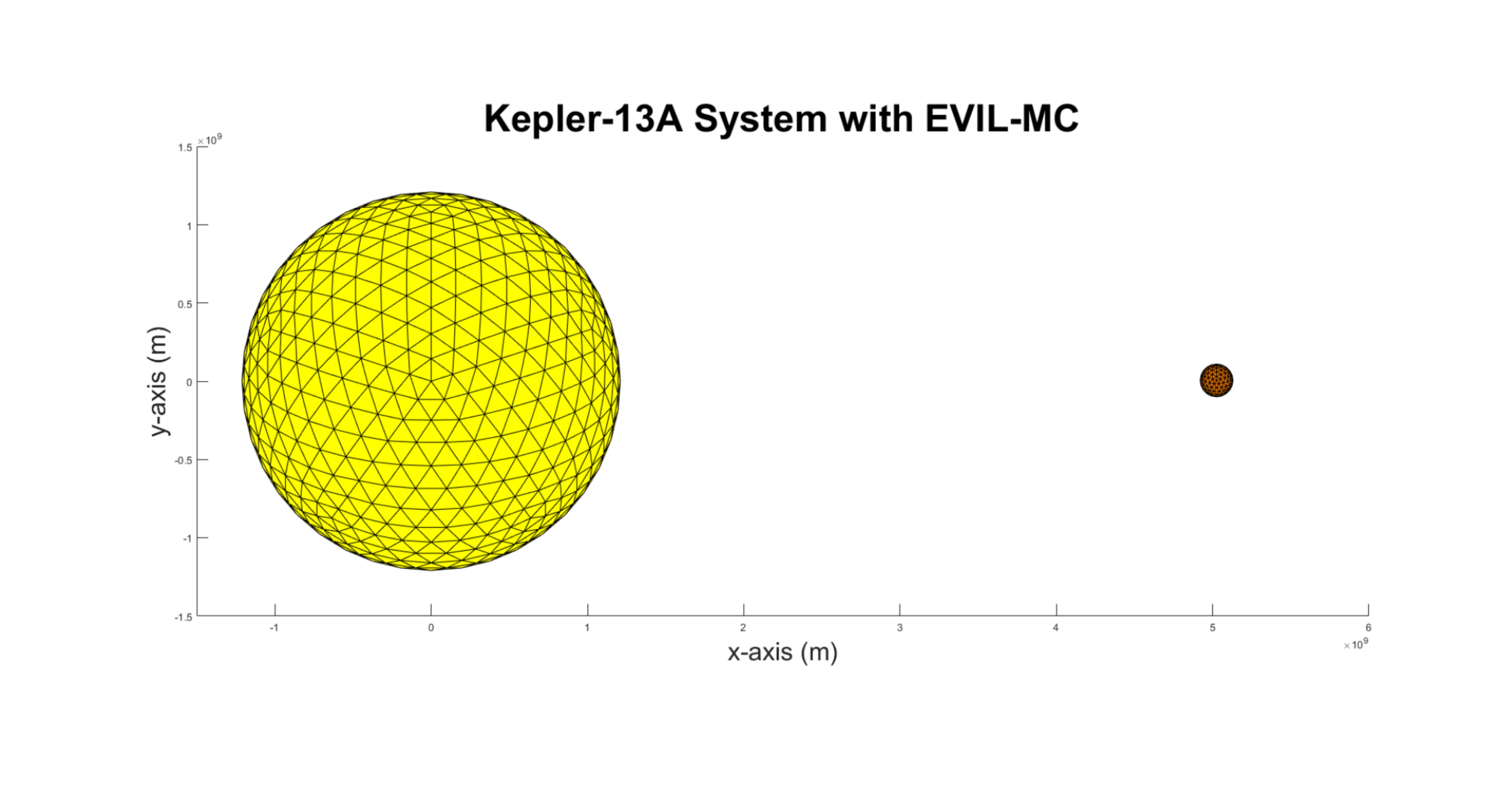}\label{Kepler13}
    \caption{A scale model looking edge-on to the Kepler-13A system using EVIL-MC for EXONEST. A geodesic sphere provides the basic structure of the star for which the vertices provide locations where the deviations to the stellar surface have been computed. The deviations are applied to the vertices to construct the ellipsoidal star. Kepler-13A~b is a massive planet, estimated to be around 6-9~$M_J$ (\cite{Shporer2011,Shporer2014,Saterne2012}), with an orbital period of 1.763 days. A large planet orbiting closely to the star should provide significant ellipsoidal variation; however, while the effect may be detectable to Kepler, the distortions on the star are not easily observable to the eye.}\label{fig:Kepler13Geodesic}
\end{figure*}

The EXONEST version of EVIL-MC works by constructing a geodesic dome, with $f=8$ resulting in 640 triangles, where the frequency $f$ indicates the number of times each base triangle is subdivided (\cite{MakeIcosahedron}). The geodesic dome is useful because it is constructed of a nearly uniformly distributed number of points along the sphere, which form easy-to-work with triangles. Each triangle is formed with three grid-points. Additionally, only half of the star is possible to observe at any given time; therefore, only half of the star needs to be constructed. This improves the efficiency of the computation over the original EVIL-MC code. A frequency of 8 will construct a denser distribution of points compared to the 31x31 grid over a full sphere used in the original EVIL-MC and may be generated in less time. A snapshot of Kepler-13A system from an edge-on perspective is displayed using geodesic spheres in Figure \ref{fig:Kepler13Geodesic}.

The deviations from spherical shape are computed for each grid point on the geodesic sphere using Equation \eqref{EVILMC_dr}. These deviations are added to the locations of the grid point, effectively adjusting the radial distance for the surface of the star. Limb-darkening and gravity-darkening profiles can be applied for the center of each triangle. The observed area is computed by taking the dot product between the area vector of a triangle and the line-of-sight direction ($z$-direction). The observed flux within the Kepler bandpass is computed by multiplying the blackbody function with the Kepler Response Function and observed area (\cite{KeplerHandbook}). The total flux is summed for the ellipsoidal star and a spherical reference star. 


The ellipsoidal variation observed is computed with

\begin{equation}\label{PhiEllipse_EVILMC}
\frac{F_{ellip}}{F_{S}} = \frac{\phi_{Star}}{\phi_{Sphere}}-1,
\end{equation}
where $\phi_{Sphere}$ is the flux from a spherical star and $\phi_{Star}$ is the flux from the ellipsoidal star.

Stellar atmospheres may be distorted due to rotation or tidal interaction between a companion. In 1924 von Zeipel first outlined the theory for the radiative equilibrium of rotating gaseous and stellar material and the interaction of binary stars (\cite{vonZeipel1924a, vonZeipel1924b, vonZeipel1924c}). Rotation produces an extended atmosphere perpendicular to the rotation axis while compression occurs parallel with the axis of rotation. Similarly, gravitational interactions produce an extended atmosphere bulge along a vector connecting the center of masses of two bodies. A temperature gradient occurs between the hotter poles of the star and the cooler, extended atmosphere. Photometric emission is reduced in cooler regions resulting in what's known as gravity-darkening. Gravity-darkening is most often observed in binary star systems (\cite{Djurasevic2003, Djurasevic2006, White2012}) and with rapidly rotating stars (\cite{Lara2011, Szabo2012, Masuda2015, Howarth2017}). Gravity-darkening is modeled in the original EVIL-MC algorithm by looking at changes to the gravity vector on the stellar surface. This effect has not been implemented in the EVIL-MC algorithm implemented in EXONEST because gravity-darkening could not be effectively accommodated by the other ellipsoidal models. As stellar rotation is not included in this analysis gravity-darkening will provide the smallest contribution to the Kepler-13 light curve and therefore has lower priority.


While computationally intensive, this model should improve upon the accuracy provided by both BEER and the Kane and Gelino models by directly accounting for limb-darkening and gravity-darkening. Additionally, none of the trigonometric models are able to model the photometric effects due to stellar rotation.

\subsection{All Models Comparison}

\begin{figure*}[htb!]
\centering
    \includegraphics[width=0.95\textwidth]{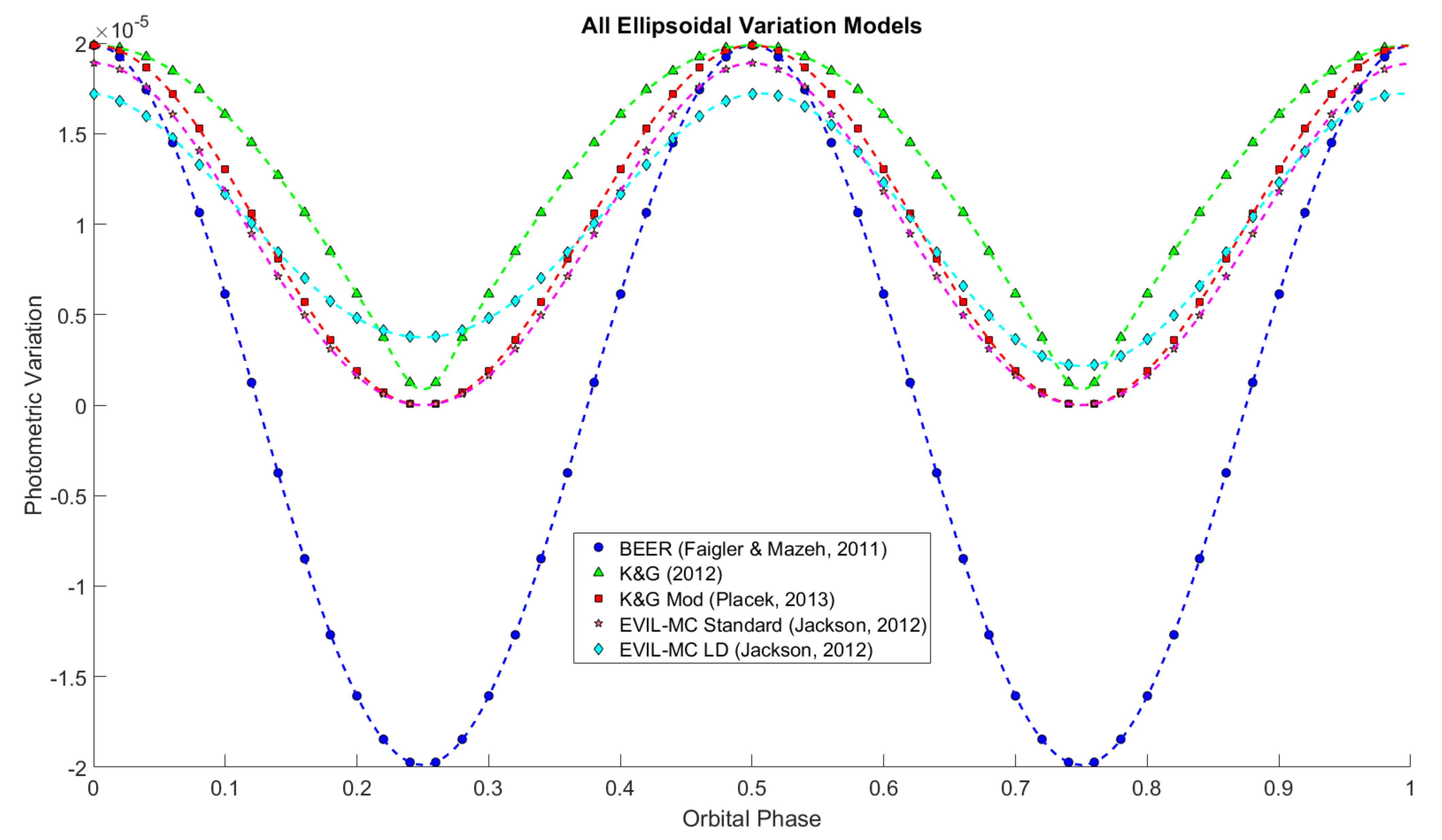}
        \caption{Shown here is each of the ellipsoidal models plotted over one orbital period. The ellipsoidal variations are based on a 1~M$_J$ mass planet in a circular orbit around a 1~M$_{\odot}$ star with an orbital period of 1 day. The EVIL-MC Standard model plotted in magenta stars represents only distortions due to the stellar shape. The EVIL-MC LD model applies a quadratic limb-darkening mask.}\label{fig:All_Models_Plot}
\end{figure*}

Each model exhibits the general characteristics of having two maxima per orbit, located around one quarter and three-quarters orbital phase, and has a maximum amplitude roughly described by \cite{Loeb2003}. All models are plotted in Figure \ref{fig:All_Models_Plot}. Included are two versions of EVIL-MC: one shown in magenta stars, which models only the star shape, and a second in cyan diamonds, which applies a quadratic limb-darkening mask. Also plotted are the trigonometric models. The modified version of Kane \& Gelino (red squares) seems to best approximate the version of EVIL-MC, which only models the stellar shape (magenta stars).
%

\subsection{Priors}

We assign uniform probabilities to each parameter within the range defined by the constraints or our search area, which are listed in Table \ref{Table_Priors}. An example parameter is orbital inclination. To select a uniform prior on the region, we select from a uniform distribution over a sphere. Impossibilities are also ruled out by the priors. The priors for planetary radius and planetary mass have been chosen to span the range from below the detectability threshold and into the brown dwarf region. A posterior containing many results with masses larger than the boundary between planet and brown dwarf of around 13$~M_J$ will lead to the classification of the object as a brown dwarf rather than an exoplanet.


\begin{table}[htb!]
\begin{center}
\caption{Prior Distributions and Ranges}\label{Table_Priors}


\begin{tabular*}{\linewidth}{@{\extracolsep{\fill}}p{0.22\linewidth}p{0.22\linewidth}p{0.22\linewidth}p{0.22\linewidth}@{}}
    \hline
    Parameter & Variable & Interval & Distribution \\ \hline \hline
	Orbital Period (Days) & $T$ & [0.01, 15] & Uniform \\
	Stellar Mass & $M_S$ & ... & Known \\
	Mean Anomaly & $M_0$ & [0, 2$\pi$] & Uniform \\
	Arg. of Periastron & $\omega$ & [0,2$\pi$] & Uniform \\
	Orbital Inclination & $i$ & [0,$\frac{\pi}{2}$] & Uniform on a Sphere \\
	Planetary Radius ($R_J$) & R$_P$ & [10$^{-4}$,20] & Uniform \\
	Stellar Radius & $R_S$ & ... & Known \\
	Planetary Mass ($M_J$) & $M_P$ & [0.1, 20] & Uniform \\
	Dayside Temperature (K) & T$_d$ & [0,5000] & Uniform \\
	Nightside Temperature (K) & T$_n$ & [0,5000] & Uniform \\
	Standard Deviation of Noise (ppm) & $\sigma$ & [10$^{-6}, 10^{-4}$] & Uniform \\
	Limb-Darkening Coefficients & $\mu_1,~\mu_2$ & ... & Known \\
	    \hline
  \end{tabular*}

\end{center}

A list of the priors assigned to each parameter. Distributions may be adjusted beyond uniform, given additional information. Parameters assigned specific values within the prior are `known' and assigned the known value.
\end{table}

Some parameters like the limb-darkening and gravity-darkening coefficients were taken as known quantities and were determined by using prior knowledge of the star parameters.

\section{Application}\label{Application}




\subsection{Model Differences}

\begin{figure*}[t!]
\centering
    \includegraphics[width=1.00\textwidth]{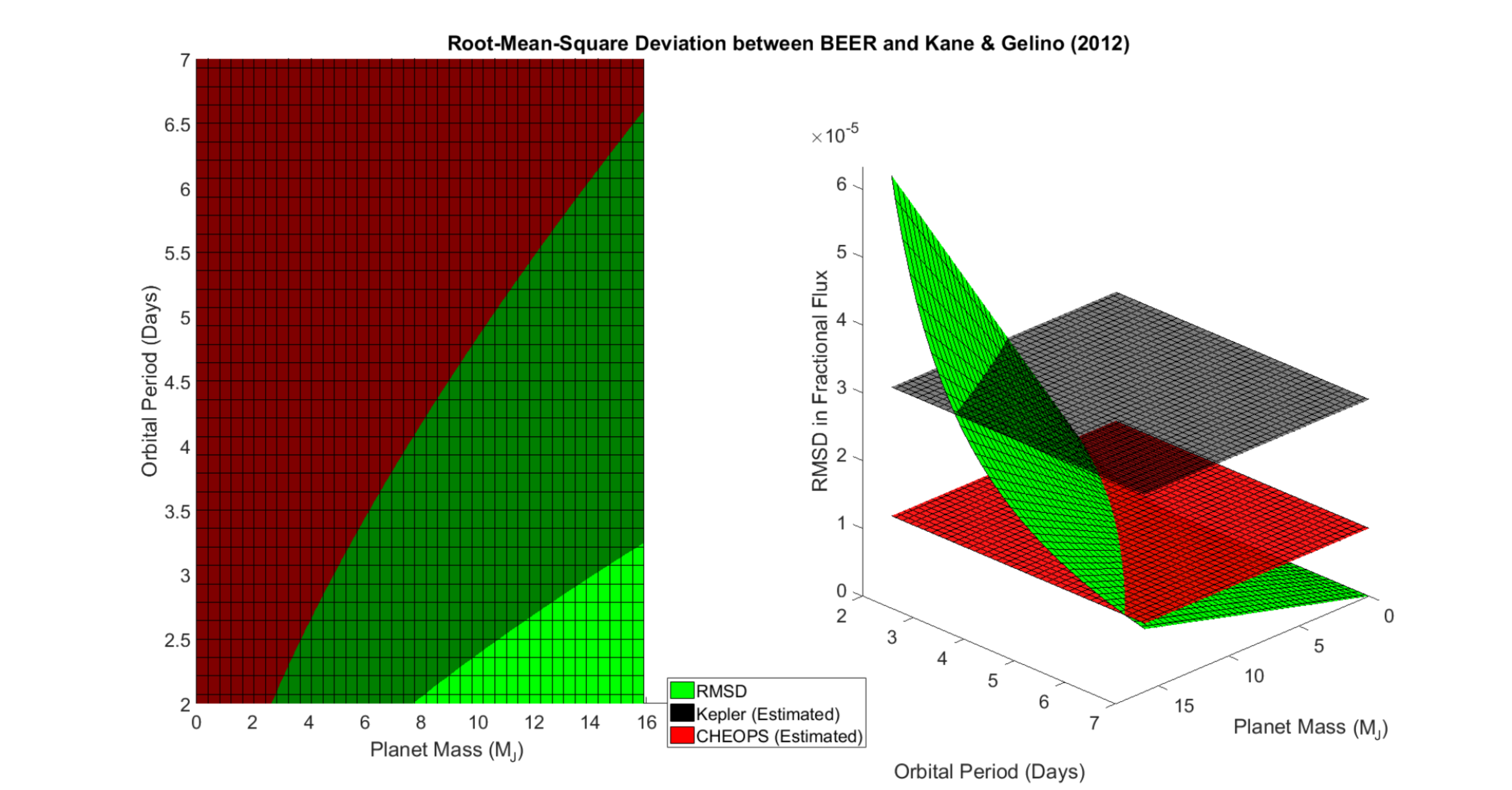}
    \caption{The RMS difference in fractional flux between the BEER and Kane \& Gelino (2012) model. The planet mass range extends between [0,16]~M$_J$ and the orbital period of [2,7]~days. The estimated Kepler precision of 30~ppm is plotted against the root-mean-square deviation between the two models as a black plane. The estimated precision from 6 hours integration time for CHEOPS is represented by the red plane. With alpha channels, this creates a bright green patch where the differences between the models should be clearly seen in Kepler data. The dark green region represents orbital periods and planetary masses for which Kepler may not be able to distinguish the models but should be detectable with optimal CHEOPS data. Red regions will likely require higher-precision (order of ppm) experiments than currently available to distinguish between the models.}\label{fig:BEER_KG2012}
\end{figure*}

\begin{figure*}[t!]
\centering
    \includegraphics[width=1.00\textwidth]{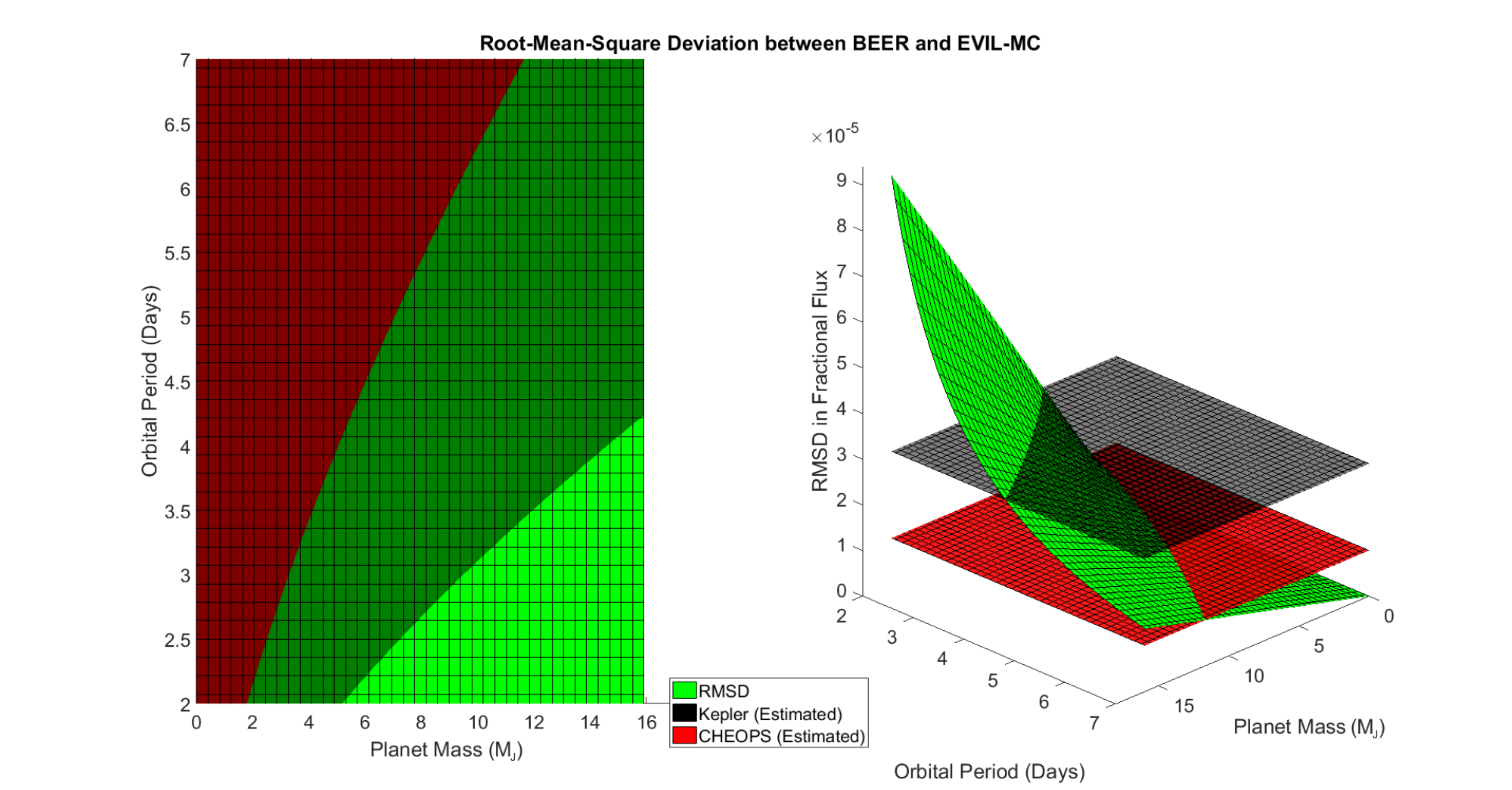}
    \caption{The RMS difference in fractional flux between the BEER and EVIL-MC model containing limb-darkening are plotted at different planet masses and orbital periods. The planet mass range extends between [0,16]~M$_J$ and the orbital period of [2,7]~days.}\label{fig:BEER_EVILMC4}
\end{figure*}

Examination of the root mean square (RMS) deviation between the models may reveal the extent to which the models may be distinguished. The ellipsoidal variation light curve for each model was generated at a given set of orbital parameters spanning planetary masses between 0 and 15~$M_{J}$ and orbital periods between 2 and 8~days. Larger differences between the models should occur when the amplitude of the ellipsoidal variations is increased. This occurs at closer orbits (shorter orbital periods) and larger planet masses. While the RMS difference between the models may be within the noise of the telescope, measurements over multiple orbits may still contain enough information to distinguish the models. The Kepler telescope has an estimated precision of 30~ppm (black plane on Figures \ref{fig:BEER_KG2012} and \ref{fig:BEER_EVILMC4}) for a 10~mag star (\cite{KeplerDataCharacteristics}). The CHEOPS mission (red plane) will reduce the precision level down to around 10~ppm (\cite{CheopsRedbook}) with 6 hr of integration time. For \textit{CHEOPS}, the ellipsoidal models should be clearly distinguishable in most hot Jupiter systems. 

The largest deviation between the models occurs between the BEER and EVIL-MC (with limb-darkening) models, which can be seen in Figure \ref{fig:All_Models_Plot}. Because the BEER model estimates the variation using a sinusoid, the model predicts a ``negative'' flux when the smallest portion of the star is being observed. EVIL-MC (with limb-darkening) estimates the flux to be positive at this point.  This occurs at orbital phase angles of 0.25 and 0.75 in Figure \ref{fig:All_Models_Plot}. This is the largest difference between all of the models at any point along the orbit.

\subsubsection{Synthetic Kepler Data}

The ellipsoidal models were compared using synthetic data generated by adding Gaussian noise to data generated by using each of the models in turn. Models with noise at levels 0, 10, and 30~ppm were compared, which includes the estimated uncertainty of the future mission, \textit{CHEOPS}, and current mission, Kepler. Analysis on synthetic data with Gaussian noise added up to 30~ppm was able to determine the original model for ellipsoidal variation. Increasing the noise reduced the distinguishability of the models. 



\begin{figure*}[htb!]
\centering
    \includegraphics[width=0.95\textwidth]{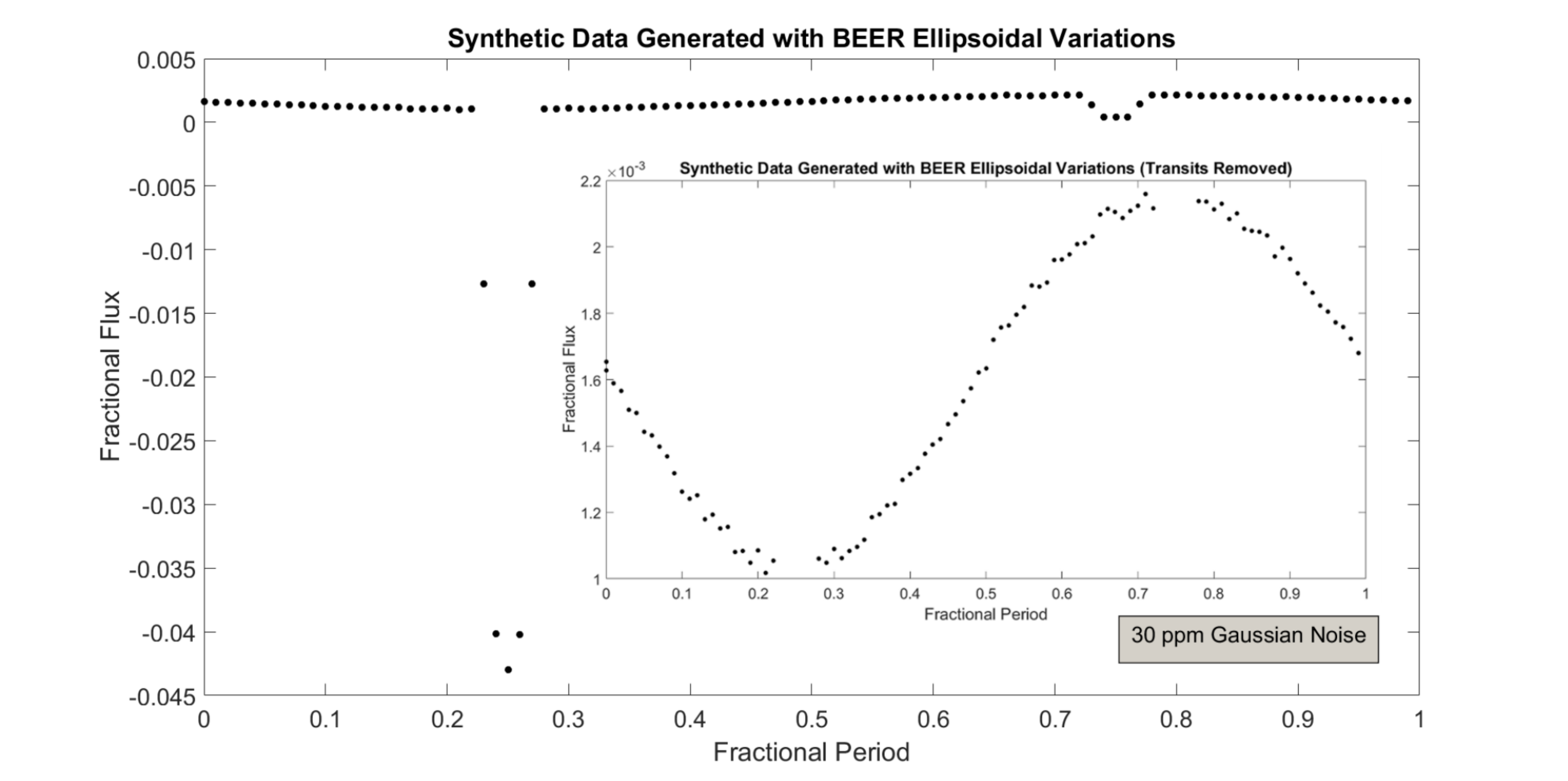}

    \caption{A plot showing the synthetic light curve data generated using the BEER model for ellipsoidal variations. Gaussian noise has been added with a maximum amplitude of 30~ppm.  A plot showing the photometric variations has also been included by removing data points within the primary and secondary transits. The effect from thermal emission dominates the photometric variations. The thermal variations are of order $10^{-3}$ while the ellipsoidal variations and Doppler boosting effects are on the order of $10^{-5}$ in the fractional flux.}    \label{fig:Synthetic_BEER}
\end{figure*}


%

\begin{table*}[t!]

\caption{Model Testing Synthetic Data No Noise} \label{Table_Synthetic_Results_0ppm}
\begin{center}
\resizebox{0.95\textwidth}{!}{\begin{minipage}{\textwidth}
  \begin{tabular} {| c || c || c || c || c || c |}
    \hline
\backslashbox{Model}{Data} & BEER & Kane \& Gelino (2012) & Kane \& Gelino (Mod) & No EV & EVIL-MC \\ \hline\hline
BEER & \textbf{1223.25 $\pm$ 1.00} & 1020.84 $\pm$ 0.89 & 1044.640 $\pm$ 0.90 & 979.25 $\pm$ 0.86 & 1020.81 $\pm$  0.89 \\ \hline
Kane \& Gelino (2012) & 967.93 $\pm$ 0.86 & \textbf{1223.46 $\pm$ 1.00} & 1103.32 $\pm$ 0.93 & \textbf{1012.92 $\pm$ 0.89} & 885.85 $\pm$ 0.93 \\ \hline
Kane \& Gelino (Mod) & 1001.64 $\pm$ 0.78 & 1114.35 $\pm$ 0.94 & \textbf{1223.92 $\pm$ 1.00} & 920.86 $\pm$ 0.87 & \textbf{1098.96 $\pm$ 1.01} \\ \hline
No EV & 903.51 $\pm$ 0.81 & 991.61 $\pm$ 0.87 & 968.11 $\pm$ 0.85 & 995.07 $\pm$ 0.94 & 984.54 $\pm$ 0.91 \\ \hline
\end{tabular}
\end{minipage} }
\end{center}
The data used in this test contained no added noise and included other photometric effects present in light curve data (see Section \ref{Sec:OtherPhotoEffects}). Each row represents the evidence value for the model. Each column indicates which model was used to generate the synthetic data. Bold indicates the preferred model for each trial. The case ``No EV'' contained no ellipsoidal variations added to the light curve however; all other photometric effects are modeled. EVIL-MC was not used in the model testing because testing the direct computation model on sinusoidal data would not provide much benefit.
\end{table*}

\begin{table*}[t!]

\caption{Model Testing Synthetic Data 10 ppm Noise}\label{Table_Synthetic_Results_10ppm}
\begin{center}
\resizebox{0.95\textwidth}{!}{\begin{minipage}{\textwidth}
  \begin{tabular} {| c || c || c || c || c || c |}
    \hline
\backslashbox{Model}{Data} & BEER & Kane \& Gelino (2012) & Kane \& Gelino (Mod) & No EV & EVIL-MC \\ \hline\hline
BEER & \textbf{1011.74 $\pm$ 0.88} & 991.77 $\pm$ 0.87 & 999.40 $\pm$ 0.87 & 959.99 $\pm$ 0.85 & 970.89 $\pm$ 0.94 \\ \hline
Kane \& Gelino (2012) & 953.03 $\pm$ 0.84 & \textbf{1015.77 $\pm$ 0.88} & 1008.63 $\pm$ 0.87 & 983.49 $\pm$ 0.86 & \textbf{1014.18 $\pm$ 0.87} \\ \hline
Kane \& Gelino (Mod) & 973.71 $\pm$ 0.85 & \textbf{1012.92 $\pm$ 0.54} & \textbf{1020.56 $\pm$ 0.88} & 974.48 $\pm$ 0.85 & 951.80 $\pm$ 0.91  \\ \hline
No EV & 899.74 $\pm$ 0.81 & 970.00 $\pm$ 0.84 & 952.96 $\pm$ 0.83 & \textbf{1015.02 $\pm$ 0.88} & 973.53 $\pm$ 0.85 \\ \hline
\end{tabular}
\end{minipage} }
\end{center}
The data used in this test contained 10~ppm of Gaussian noise, which equivalent to an estimated 6~hr integration time of the \textit{CHEOPS} telescope. The data also included other photometric effects present in light curve data (see Section \ref{Sec:OtherPhotoEffects}).
\end{table*}


\begin{table*}[t!]
\caption{Model Testing Synthetic Data 30 ppm Noise}\label{Table_Synthetic_Results_30ppm}

\begin{center}
\resizebox{0.95\textwidth}{!}{\begin{minipage}{\textwidth}
  \begin{tabular} {| c || c || c || c || c || c |}
    \hline
\backslashbox{Model}{Data} & BEER & Kane \& Gelino (2012) & Kane \& Gelino (Mod) & No EV & EVIL-MC \\ \hline\hline
BEER & \textbf{911.96 $\pm$ 0.82} & 914.04 $\pm$ 0.82 & 908.78 $\pm$ 0.82 & 915.65 $\pm$ 0.81 & \textbf{912.85 $\pm$ 0.82} \\ \hline
Kane \& Gelino (2012) & 896.94 $\pm$ 0.80 & \textbf{918.30 $\pm$ 0.81} & \textbf{913.13 $\pm$ 0.81} & \textbf{920.21 $\pm$ 0.82} & \textbf{915.24 $\pm$ 0.83} \\ \hline
Kane \& Gelino (Mod) & 902.57 $\pm$ 0.81 & \textbf{918.88 $\pm$ 0.81} & \textbf{912.94 $\pm$ 0.81} & 918.69 $\pm$ 0.82 & \textbf{917.40 $\pm$ 0.82} \\ \hline
No EV & 875.28 $\pm$ 0.78 & 907.98 $\pm$ 0.81 & 899.50 $\pm$ 0.79 & \textbf{925.10 $\pm$ 0.81} & 906.30 $\pm$ 0.83 \\ \hline
\end{tabular}
\end{minipage} }
\end{center}
The data used in this test contained 30~ppm of Gaussian noise, which is equivalent to the estimated raw photometric uncertainty in Kepler Long Cadence data for magnitude 10 stars. The data also included other photometric effects present in light curve data (see Section \ref{Sec:OtherPhotoEffects}).
\end{table*}

 In every case, with the exception of the data with no noise and no added variation, the constructing model had the largest evidence. The $logZ$ evidences are presented in Tables \ref{Table_Synthetic_Results_0ppm}-\ref{Table_Synthetic_Results_30ppm} with the most preferred model indicated in bold. If two models were nearly indistinguishable $(logZ < 5)$, both models are in bold. The Kane \& Gelino (2012) model and the Kane \& Gelino (Modified) model were difficult to distinguish with increasing noise level. The Kane \& Gelino (2012) model and the Kane \& Gelino (Modified) models are similar in structure; that is, their residuals are small, which can be seen in Figure \ref{fig:All_Models_Plot}. Generating data with either model will produce a similar light curve, thus contributing to their indistinguishably. Both models should perform similarly well for noisy data if the data was constructed using either the Kane \& Gelino (2012) model or the Kane \& Gelino (Modified) model.
 

%

\subsection{Modeling the Kepler-13A System}

Kepler-13A~b was detected by \cite{Shporer2011} via only photometric effects using the BEER algorithm. The Kepler-13 system is currently believed to contain four major bodies: an A-type star with a companion Jupiter-like exoplanet and another A-type star with a stellar companion of spectral type K or M (\cite{Mislis2012, Saterne2012, Shporer2014}). Stellar rotation has been measured on Kepler-13A and was used to update the stellar radius by \cite{Howarth2017}. \cite{Szabo2012} report a spin-orbit resonance, transit duration variations, and possible secular perturbations for Kepler-13A. \cite{Masuda2015} performed spin-orbit measurements for the Kepler-13A system. Geometric albedo, bond albedo, and temperature estimates of Kepler-13A~b were found by \cite{Placek2013, PlacekThesis, Angerhausen2015}. The Kepler-13 system was chosen for model testing using EXONEST because the host star is relatively quiet and Kepler-13A~b is a large planet with a short orbital period with confirmed detections using both radial velocity and transit methods. This provides the best conditions to observe ellipsoidal variations. The Kepler-13 system has an apparent magnitude in the Kepler bandpass of around magnitude 10. The estimated precision of Kepler long cadence data is around 29~ppm (\cite{KeplerPrecision}). Effects smaller than $\approx$ 29~ppm are increasingly difficult to detect. The parameter values of for the Kepler-13A system are displayed in Table \ref{Table_Kepler13}.

\begin{table}[h!]
\caption{Kepler-13A System Priors}\label{Table_Kepler13}
\begin{center}
\begin{tabular*}{\linewidth}{@{\extracolsep{\fill}}p{0.3\linewidth}p{0.3\linewidth}p{0.3\linewidth}@{}}
    \hline
    Parameter (Units) & Symbol & Value  \\ \hline \hline
    Stellar Mass (M$_{\astrosun}$) & M$_{\astrosun}$ & $2.466^{+0.447}_{-0.7250}$ \\ \hline
    Stellar Radius (R$_{\astrosun}$) & R$_{\astrosun}$ & $3.031^{+1.198}_{-0.944}$ \\ \hline
    Stellar Effective Temperature (K) & $T_{Eff}$ & $9107^{+257}_{-425}$ \\ \hline
    Orbital Period (Days) & $T$ &1.76358756 $\pm$3.2E-08 \\ \hline
    Metallicity (dex) & $Z$ & 0.070$^{+0.14}_{-0.65}$ \\ \hline
    Limb-darkening Coefficient & $u_1$ & 0.2278 \\ \hline
    Limb-darkening Coefficient & $u_2$ & 0.2884 \\ \hline
    Gravity-darkening Coefficient & g & 0.5319 \\ \hline

  \end{tabular*}
\end{center}
Table containing the values assumed to be known in the trials for the Kepler-13A system. Values are recorded from the NASA Exoplanet Archive. The system priors are assumed to be precisely known with the reported values. The uncertainties of these parameters are ignored in this analysis. The limb-darkening coefficients were determined by interpolation from coefficients computed by \cite{Claret2011}. Kepler-13~B and its stellar companion are not modeled.

\end{table}



\subsection{Other Photometric Effects Modeled}\label{Sec:OtherPhotoEffects}

In addition to ellipsoidal variations, the other effects modeled are transits, thermal emission, and Doppler boosting.

\subsubsection{Transits}

In EXONEST, the depth of the primary transit (when the planet blocks out light from the star) is modeled using the cross-sectional areas of the planet and the star. The ratio was simplified by \cite{Seager2003} to

\begin{equation}
\delta F_P = \Big( \frac{R_P}{R_S}\Big)^2,
\end{equation}
where $R_P$ is the radius of the planet and $R_S$ is the radius of the star.

A secondary eclipse can occur when light from the planet is obscured by the star. The depth of the secondary transit is modeled using

\begin{equation}
\delta F_P = \frac{T_{d}}{T_{eff}} \Big( \frac{R_P}{R_S} \Big)^2,
\end{equation}
where $T_d$ and $T_{eff}$ are the day-side temperature of the planet and effective temperature of the star, respectively. \cite{Mandel2002} provided analytic solutions to transit light curves for quadratic limb-darkening law and a non-linear limb-darkening law. Currently, EXONEST implements the quadratic limb-darkening law.

\subsubsection{Thermal Emission}

Planets that are close-in to their stars will likely be intensely hot, potentially in the thousands of degrees Kelvin. The Kepler Space Telescope observes wavelengths from around 420~nm to 900~nm. As such, the blackbody radiation from the planet is detectable within the bandpass of Kepler.



The thermal contribution is broken up between day-side and night-side components. The two components are shifted by half an orbit from each other assuming that exactly half of the planet is illuminated by the star. The day-side contribution is given by:

\begin{equation}
\frac{F_{Th,d}(t)}{F_S} = \frac{1}{2} \Big(1 + \cos \theta(t) \Big) \Bigg(\frac{R_P}{R_S}\Bigg)^2 \frac{\int B(T_d,\lambda)K(\lambda) d\lambda}{\int B(T_{eff}, \lambda)K(\lambda) d\lambda},
\end{equation}
where $B(T_d, \lambda$) is the wavelength-dependent blackbody radiation at the day-side temperature of the planet, $B(T_{eff},\lambda$) is the blackbody radiation from the effective temperature of the star at wavelength $\lambda$, and $K(\lambda$) represents the wavelength-dependent Kepler response function. Similarly, the night-side contribution is given by

\begin{equation}
\frac{F_{Th,n}(t)}{F_S} = \frac{1}{2} \Big(1 + \cos \big( \theta(t) - \pi \big) \Big) \Bigg(\frac{R_P}{R_S}\Bigg)^2 \frac{\int B(T_n, \lambda)K(\lambda) d\lambda}{\int B(T_{eff}, \lambda)K(\lambda) d\lambda},
\end{equation}
where $T_{n}$ is the night-side temperature of the planet (\cite{Placek2013,PlacekThesis}).

Reflected light also appears in the light curve with a period equal to the orbital period of the planet and is maximum at the full phase. For circular orbits, the reflected light and thermal light cannot be distinguished. Trying to model both effects simultaneously leads to degeneracies in the model. To prevent unnecessary degeneracies, reflected light was not included. The planetary emission component of the light curve is modeled entirely as thermal emission. Thermal emission was shown to be more significant than reflected light in the Kepler-13 light curve by \cite{Placek2013,PlacekThesis}.

Each of the models for thermal emission and reflected light assume incoming parallel rays from the host star. \cite{Knuth2017} explores more detailed models accounting for non-parallel incident rays that are required for large stars and planets with extremely close-in orbits.

\subsubsection{Doppler Boosting}

The star and planet(s) orbit the barycenter of the system. This leads to a photometric variation arising from the relativistic effect of stellar aberration. As the star moves toward the observer, there is an increase in observed flux. As the star recedes from the observer, the observed flux is reduced.

The radial velocity (along $\hat{z}$) may be represented using the orbital anomalies as

\begin{equation}
V_z(t) = K\Big( \cos \big(\nu(t) + \omega \big) + e\cos \omega\Big),
\end{equation}
where $K$ is defined as the radial velocity semi-amplitude

\begin{equation}
K = \Big(\frac{2\pi G}{T}\Big)^{1/3} \frac{m_P \sin i}{m_S^{2/3} \sqrt{1-e^2}}.
\end{equation}

The normalized line-of-sight velocity, $\beta_r(t)$, for the star is

\begin{equation}
\beta_r(t) = \frac{V_z(t)}{c}.
\end{equation}

The effect is modeled by
\begin{equation}
\frac{F_{boost}(t)}{F_S} = 4 \beta_r(t),
\end{equation}
and was originally derived in \cite{Rybicki1974} and reduced to the non-relativistic limit by \cite{Loeb2003}.

\subsection{Analysis of Ellipsoidal Variations in Kepler-13}

EXONEST was run using each model for ellipsoidal variations using 75 samples within the MultiNest algorithm and the known values from Table \ref{Table_Kepler13}. For stopping criterion, the tolerance in MultiNest was set to 0.01. For the Kepler-13 system, EVIL-MC (with quadratic limb-darkening) provided the best evidence with an average $LogZ = 14,258.21 \pm 0.65$. Model comparison may be done using the Bayes' factor, which is constructed by taking the ratio of the evidences (or differences of the logarithmic evidences). The trigonometric models BEER and Kane \& Gelino (Modified) were nearly indistinguishable from EVIL-MC with Bayes' factors $e^{2.32}~(\approx10.2)$ and $e^{1.91}~(\approx6.8)$ respectively. All three models were preferred over the Kane \& Gelino (2012) model with a Bayes' Factor of around $e^{3.0}$-$e^{5.3}~(\approx20.1-200.3)$. 

\begin{figure*}[htb!]
\centering
    \includegraphics[width=0.95\textwidth]{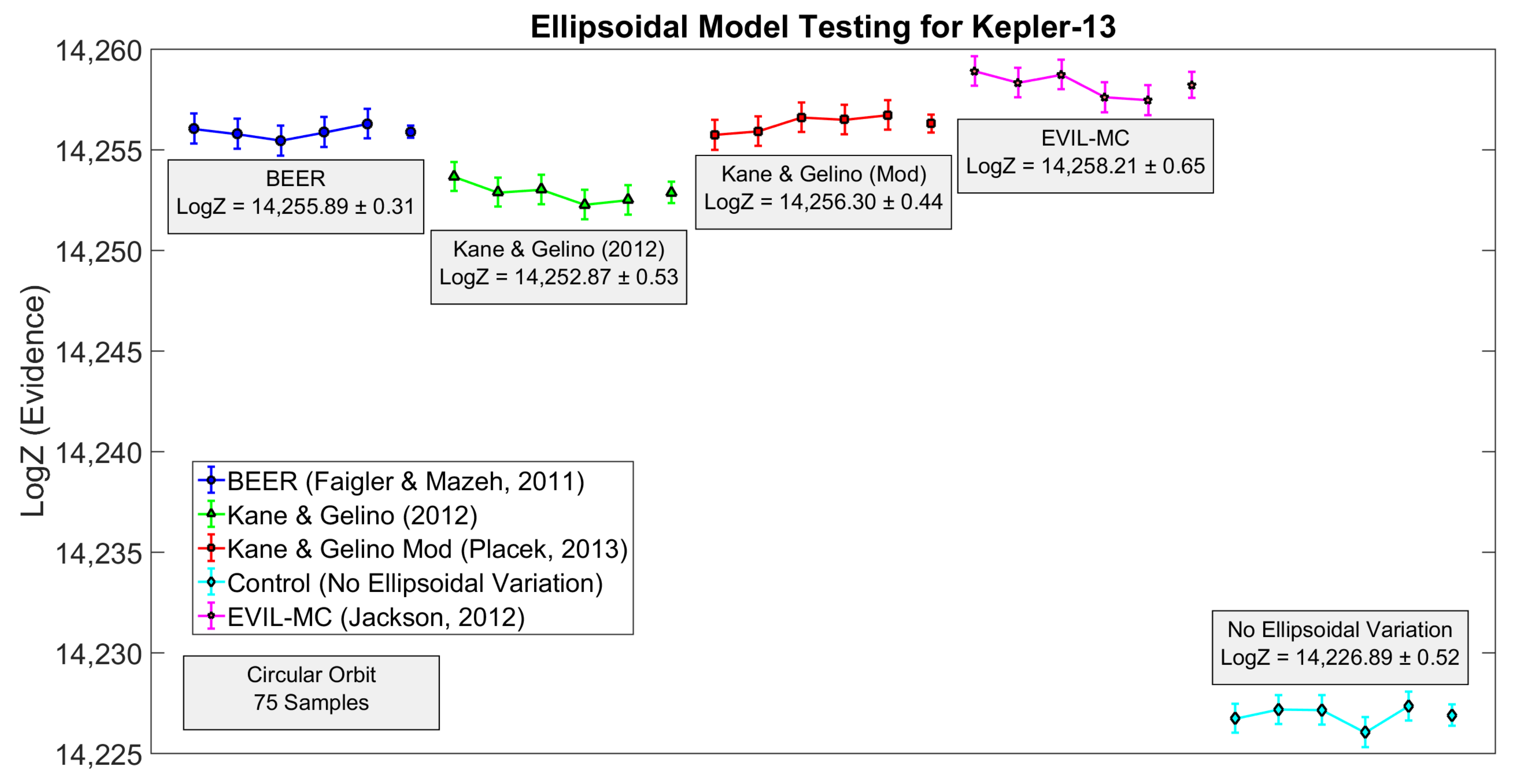}

    \caption{Plot displaying the evidence values for five runs of each of the models. A model where the ellipsoidal variation was set to zero is also displayed. EVIL-MC performed the best and is slightly preferred over the modified Kane \& Gelino model and BEER model. The modified Kane \& Gelino model performed the best of the trigonometric models but was nearly indistinguishable from the BEER model. Both were slightly preferred over Kane \& Gelino (2012). All of the ellipsoidal variation models were significantly preferred over no variation, which indicates ellipsoidal variations are present in Kepler-13 data. All models tested included the other photometric effects mentioned in Section \ref{Sec:OtherPhotoEffects}}    \label{fig:Kepler13EvidencesCircular}
\end{figure*}

All models for ellipsoidal variations were significantly preferred over a model containing no variation. Therefore, ellipsoidal variations are detectable within the Kepler-13 system. The evidence for each run is plotted in Figure \ref{fig:Kepler13EvidencesCircular}.

All of the models agreed on orbital parameters other than the mass of the planet. The BEER model estimated the planet mass at around 1~M$_J$. This is significantly less than the masses estimated by the other models (see Table \ref{Kepler13_Posteriors}). The mean value of the variation using the BEER model is zero. For both Kane \& Gelino models, the effect is always greater than or equal to zero. This means that the BEER model only requires a planet of around half the mass of the Kane \& Gelino models to represent the same amplitude (See Figure \ref{fig:All_Models_Plot}). This is confirmed in the posteriors. 





\begin{figure*}[t!]
\centering
    \includegraphics[width=0.95\textwidth]{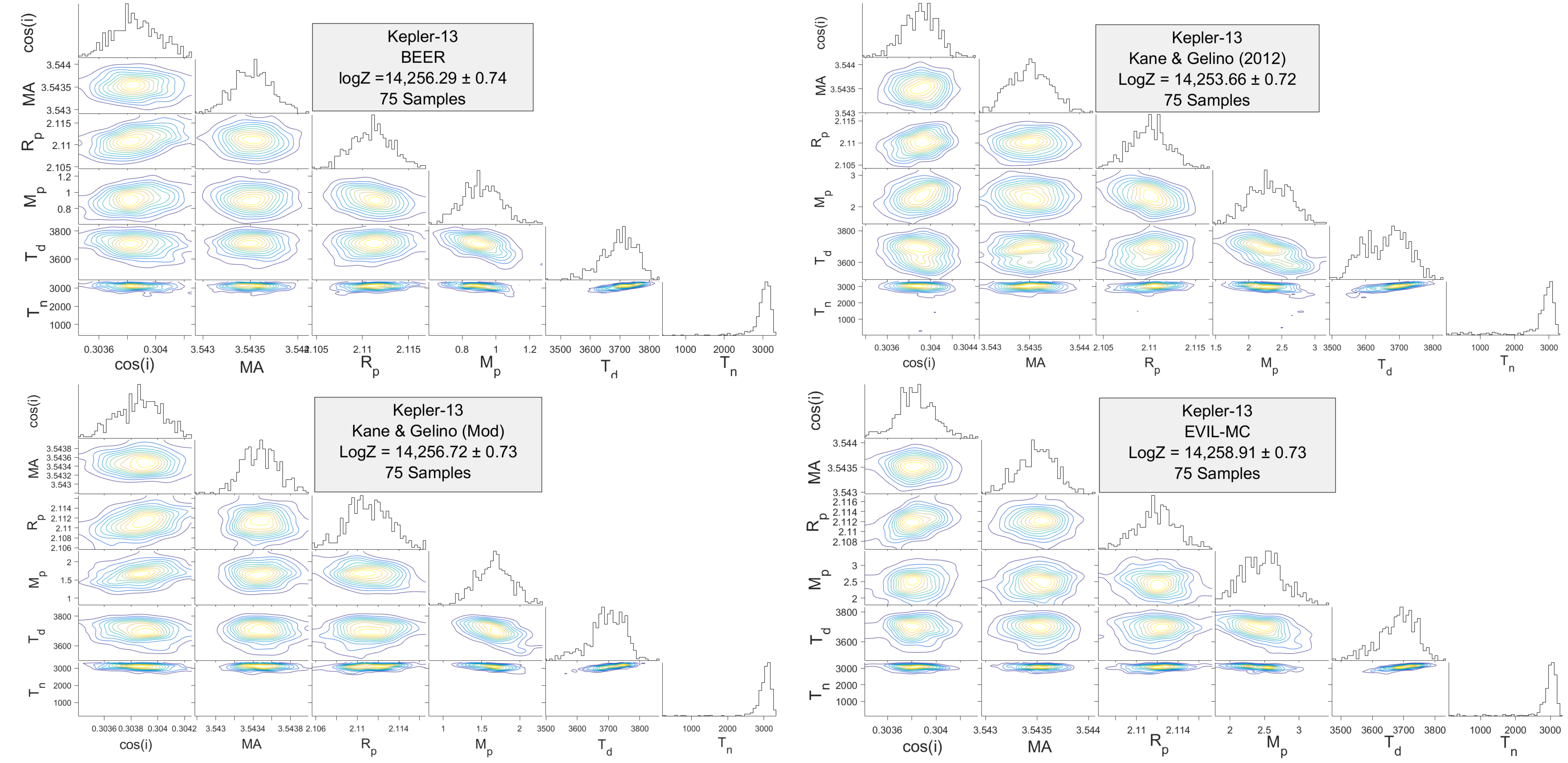}
	    \caption{Corner plot displaying the posterior of Kepler-13. Shown is $\cos(i)$, the cosine of the inclination of the planetary orbit, the mean anomoly (MA), planetary radius ($R_p$), planetary mass ($M_p$), and the day-side and night-side planet temperatures $T_d$ and $T_n$ respectively. The posterior distributions for all parameters other than planetary mass were nearly identical. The posterior means are displayed in Table \ref{Kepler13_Posteriors}. The BEER model produced a concentrated posterior with a width of around half a Jupiter-mass. The Kane \& Gelino (2012) model, Kane \& Gelino (Modified) model, and EVIL-MC produced planet-mass posteriors with a broader region of high probability spanning a range of around 1~$M_J$. The parameter $\cos(i)$ is used over the regular inclination because it is easier to incorporate into nested sampling priors. The inverse cosine function returns the traditional inclination.}\label{fig:CornerPlot}
\end{figure*}

The modified Kane \& Gelino model provided a posterior distribution nearly identical to the distribution generated using EVIL-MC. The posteriors for each model may be seen in Figure \ref{fig:CornerPlot}. All of the parameter correlations are illustrated. The posteriors are generally well-behaved for Kepler-13 in all of the models. The planetary mass has proportionately larger uncertainties than do the other parameters. This is likely due to the noise within the dataset blurring some of the photometric variations.

An additional factor to be considered in model selection is the computational speed of the model. All computations were done on an Intel(R) Xeon(R) E5-1603 @ 2.8~GHz with 8~GB RAM. The modified Kane \& Gelino model (and the other trigonometric models) will complete a full nested sampling run in minutes to approximately an hour. EVIL-MC took about a week on average to complete. The Modified Kane \& Gelino model is numerically the most similar to EVIL-MC; therefore, the Modified Kane \& Gelino model may provide a computationally fast approximation for EVIL-MC.

\begin{table*}[htb!]
\caption{Posterior Estimates for the Kepler-13Ab}\label{Kepler13_Posteriors} 
\begin{center}
  \begin{tabular}{ | c | c | c | c | c |}
    \hline
    Parameter & BEER & Kane \& Gelino (2012) & Kane \& Gelino (Modified) & EVIL-MC \\ \hline \hline
   $LogZ$ & 14,256.29 $\pm$ 0.74 & 14,253.66 $\pm$ 0.72 & 14,256.71 $\pm$ 0.73  & \textbf{14,258.91 $\pm$ 0.73}\\ \hline
    $\cos I$ & 0.30386 $\pm$ 1.6E-4 & 0.30390 $\pm$ 1.6E-4 & 0.30386 $\pm$ 1.7E-4 & 0.30383 $\pm$ 1.6E-4 \\ \hline
    Mean Anomaly &3.54352 $\pm$ 2.0E-4  & 3.54351 $\pm$ 2.0E-4 & 3.54348 $\pm$ 2.0E-4 & 3.54350 $\pm$ 2.0E-4 \\ \hline
    Planet Radius (R$_J$) & 2.1111 $\pm$ 2.2E-3 & 2.1100 $\pm$ 2.1E-3 & 2.1111 $\pm$ 2.1E-3& 2.1119 $\pm$ 2.0E-3 \\ \hline
    Planet Mass (M$_J$) & 0.91 $\pm$ 0.11 & 2.32 $\pm$ 0.31 & 1.66 $\pm$ 0.23 & 2.47 $\pm$ 0.30 \\ \hline
    Day-side Temp. (K) & 3696 $\pm$ 61 & 3661 $\pm$ 64 & 3695 $\pm$ 58 & 3680 $\pm$ 62 \\ \hline
    Night-side Temp (K) & 2973 $\pm$ 337 & 2710 $\pm$ 64 & 2969 $\pm$ 393 & 2887 $\pm$ 498 \\ \hline
  \end{tabular}
\end{center}

\end{table*}

\section{Future Work}\label{FutureWork}

\subsection{Additional Models}
Ellipsoidal variations are modeled in binary star systems. One of the traditionally used models is the Wilson–Devinney (WD) code by \cite{Wilson1971}, which is currently in use with PHysics Of Eclipsing BinariEs (PHOEBE) (\cite{Prsa2005, Prsa2016}). The WD model is written in FORTRAN, so some modifications may be required for compatibility with EXONEST. The model has had relatively frequent updates since 1971, the most recent occurring in 2007. WD works by following the classical Roche model of binaries in synchronous rotation. This should produce a result similar to EVIL-MC, but may be more accurate for low-mass stars. Similarly to EVIL-MC, the WD code can account for stellar rotation and directly models limb-darkening and gravity-darkening.

For circular orbits, the flux from reflected light and thermal emission cannot be disentangled. In general, for hot Jupiters, the thermal emission is usually the dominant component. However, this need not always be the case (\cite{Placek2014, PlacekThesis}). Reflected light could be modeled instead of thermal emission. Additionally, both the reflected light and thermal emission may be combined into one effect that represents the total flux from the planet. A joint effect will prevent prediction of temperature and albedo for the planet but still be able to provide the photometric variation presented by both effects. Alternatively, it may be more beneficial to model a single effect (either thermal emission or reflected light) and use the estimated value as a maximum estimated value of the planetary temperature or albedo. The true value is likely to be less than that estimated using such methods.

\subsection{Computational Efficiency}

Model testing in the Bayesian framework using the evidence does not take into account the effort required to compute a model. Currently, the trigonometric models are quick to compute, each taking only milliseconds to run. EVIL-MC is a much more computationally intense model, and one iteration takes at least a few thousand times longer to run (on the order of seconds). Over many tens of thousands to hundreds of thousands of iterations, the difference in computation time becomes significant, taking days instead of minutes to complete. This corresponds to significant difficulty when using an algorithm like nested sampling. A run of EXONEST on an Intel(R) Xeon CPU E5-1603 @ 2.8~GHz with 8~GB RAM using one of the trigonometric models for a circular orbit can take a few minutes to an hour to run on most datasets, depending on the amount of data and number of samples used to explore the space. The larger the datafile and the more samples used, the longer it takes to run the algorithm. For Kepler-13, the EVIL-MC model took on average about seven days to explore the parameter space, determine the posterior, and calculate the evidence.

In order to create a more usable model, EVIL-MC must be sped up by at least an order of magnitude. Computation time may be significantly improved by translating EXONEST and EVIL-MC into a faster, machine-native coding language. Python and C++ are potential options, with the former being a strong candidate for universal use within the scientific community and having MULTINEST already available in the language. Additionally, it is possible to compile the EVIL-MC code into C, then call the function through MATLAB. This will benefit from the structure of MATLAB while having the speed of a compiled language.

Many of the functions, primarily the computation of the flux emitted from a section of the star, will benefit from parallelization. Initial tests did not show a significant improvement in speed, but the optimization can be improved.




\section{Conclusions}\label{Conclusions}

Precise modeling of photometric variations within transit data will become increasingly important with the launch of future missions like \textit{CHEOPS} and \textit{PLATO}. One of the photometric variations, which is significant for large, close-in exoplanets, is ellipsoidal variations. Gravitational interactions between the stellar atmosphere and the planet cause the star to distort from a uniform spherical shape. Rotation will also cause the star to deviate from a spherical shape. In general, ellipsoidal variations will have two maxima per orbit of the planet, each occurring when the largest cross-sectional area of the star is being observed. This will occur roughly when the planet is at half and three-quarters phase. Ellipsoidal variations in photometric data are described using many different models (\cite{Faigler2010,KaneGelino2012,Jackson2012,Placek2013}). 


The Bayesian-based exoplanet detecting and characterizing algorithm EXONEST was used to evaluate the evidence and determine posterior distributions using each model for ellipsoidal variations. The nested sampling search algorithm MULTINEST was used as a framework within EXONEST (\cite{Feroz2008,Feroz2009,Feroz2013}). MULTINEST provides a search algorithm that efficiently samples high dimensional and multi-modal spaces while computing the evidence.

The models were first compared by determining the full photometric effect of ellipsoidal variations for each model for planet masses between 0 and 15~M$_J$ and orbital periods between 2 and 8~days. The rms between each model was computed at each orbital phase. The rms deviation between the BEER and Kane \& Gelino (2012) models is plotted in Figure \ref{fig:BEER_KG2012}, and the rms deviations between the BEER and EVIL-MC4 models are plotted in Figure \ref{fig:BEER_EVILMC4}; therefore, the largest difference between any of the models was determined to occur between the BEER  and EVIL-MC models. The differences between BEER and EVIL-MC should be detectable in exoplanets with masses larger than 6~M$_J$ within Kepler data. The differences between BEER and the other trigonometric models are more significant for more massive planets. The models may be distinguishable in optimal Kepler data with large planets ($>$8~M$_J$) with short orbital periods ($<$3~days).

Synthetic data were created using each of the models for ellipsoidal variations, including a null model in which the ellipsoidal variations were set to zero. The data were generated at levels of Gaussian noise ranging from 0 to 30~ppm. In nearly every case, the model used to generate the data was identified by a Bayes' Factor much larger than one. The Kane \& Gelino models were indistinguishable at the 30~ppm noise level, but both models were significantly preferred over the BEER model and the model with no ellipsoidal variations. The noise in the Kepler dataset is estimated to be 29~ppm for 10 mag stars (\cite{KeplerPrecision}).

The model testing was extended to a confirmed hot Jupiter exoplanet, Kepler-13A~b. EXONEST was run using each of the models, assuming a circular orbit. Beyond ellipsoidal variations, EXONEST may also model transits, reflected light and thermal emission from the planet, and Doppler boosting on the star.

EXONEST determined that, for the Kepler-13 system, EVIL-MC was the preferred model by a factor of around $13.60$, $20.49$, and $419.89$ over the Modified Kane \& Gelino, BEER, and Kane \& Gelino (2012) models, respectively.  All models of ellipsoidal variation were significantly preferred over no variations by a Bayes' factor of at least $7.2 \times 10^{10}$. This shows that ellipsoidal variations are a significant component of the observed flux from the Kepler-13 system.

The most significant difference between the models is the effect the planet mass has on the amplitude of photometric flux due to ellipsoidal variations. A planet of the same mass will produce approximately twice the total change in photometric flux within the BEER model as any of the other representations of ellipsoidal variations, as illustrated in Figure \ref{fig:All_Models_Plot}. Therefore, the BEER model will produce a significantly different estimated planet mass than the other models. Selection of the best model is critical in determining the planetary mass.


The synthetic data showed that the model used to construct the data was identifiable using the Bayesian evidence but had difficulty distinguishing the Kane \& Gelino (2012) model and the Modified Kane \& Gelino model at higher noise levels. This, combined with the results from Kepler-13 showing that the BEER and Modified Kane \& Gelino models were both slightly better represented than Kane \& Gelino (2012) in the evidence, shows that neither of the trigonometric models exactly represents the effect. It is likely that the effect is more similar to EVIL-MC, which is indicated in the slightly higher evidence for EVIL-MC in the Kepler-13 circular orbit dataset. It is likely that the differences introduced by BEER are compensated by slightly varying the parameters associated with the Doppler boosting and reflected light photometric variations.

Numerically, the Modified Kane \& Gelino model is the closest approximation and may serve as a quick alternative to the more complicated EVIL-MC model. EVIL-MC takes on the order of 1000-10000 times longer to compute than the simple trigonometric models. For this reason, the computationally intensive EVIL-MC is not suitable to be used in model testing. One would likely need to test all of the possible combinations of effects to determine which are present within the dataset. Performing dozens of combinations when a single run takes multiple weeks to compute is not efficient. Instead, the simpler trigonometric models may serve as a general representation of the effect for use in model testing, and the full version may be used when performing exoplanet characterization.

\acknowledgments
Thanks to Jennifer Carter, James Walsh, members of the Knuth Exoplanet Lab, the University at Albany Department of Physics. A.G. would also like to thank the UA Benevolent Association for supporting this work.

\software{Make Icosahedron \cite{MakeIcosahedron}, Corner Plot \cite{CornerPlot}, MULTINEST \cite{Feroz2008,Feroz2009,Feroz2013}}

\bibliography{ApJ_Ellip_Var_MT_v3}

\end{document}